\documentclass[pre,tighten]{revtex4}
\usepackage[cp850]{inputenc}
\usepackage{natbib}
\usepackage{amssymb,amsmath}
\usepackage{epsfig}
\usepackage{bm}
\usepackage{color}

\usepackage{graphicx}

\usepackage{color}

\newcommand\beq{\begin{equation}}
\newcommand\eeq{\end{equation}}
\newcommand\beqa{\begin{eqnarray}}
\newcommand\eeqa{\end{eqnarray}}
\newcommand{\dd}{\text{d}}

\newcommand{\al}{\alpha}

\begin{document}

%
\title{First-order contributions to the partial temperatures in binary granular suspensions at low density}
\author{Rub\'en G\'omez Gonz\'alez\footnote[1]{Electronic address: ruben@unex.es}}
\affiliation{Departamento de F\'{\i}sica,
Universidad de Extremadura, E-06006 Badajoz, Spain}

\author{Vicente Garz\'{o}\footnote[3]{Electronic address: vicenteg@unex.es;
URL: http://www.unex.es/eweb/fisteor/vicente/}}
\affiliation{Departamento de F\'{\i}sica and Instituto de Computaci\'on Cient\'{\i}fica Avanzada (ICCAEx), Universidad de Extremadura, E-06006 Badajoz, Spain}

\begin{abstract}
The Boltzmann kinetic equation is considered to evaluate the first-order contributions $T_i^{(1)}$ to the partial temperatures in binary granular suspensions at low density. The influence of the surrounding gas on the solid particles is modeled via a drag force proportional to the particle velocity plus a stochastic Langevin-like term. The Boltzmann equation is solved by means of the Chapman--Enskog expansion around the local version of the reference homogeneous base state. To first-order in spatial gradients, the coefficients $T_i^{(1)}$ are computed by considering the leading terms in a Sonine polynomial expansion. In addition, the influence of $T_i^{(1)}$ on the first-order contribution $\zeta^{(1)}$ to the cooling rate is also assessed. Our results show that the magnitude of both $T_i^{(1)}$ and $\zeta^{(1)}$ can be relevant for some values of the parameter space of the system.


\end{abstract}

\draft
\date{\today}
\maketitle

\section{Introduction}
In the last years, the understanding of granular matter under rapid flow conditions has raised the interest of many researchers due not only to its practical applications but also due to the fact that the understanding of its properties is really an exciting challenge. At a more fundamental level, the description of polydisperse granular mixtures (namely, the gaseous state of a mixture of smooth hard spheres with inelastic collisions) has been focused on the modification of the kinetic theory of molecular gases to properly adapt it to the dissipative character of collisions among particles. Although the influence of inelasticity is reflected in all the transport coefficients, the main new feature (as compared with ordinary or molecular fluids) in the hydrodynamic equations is the presence of the so-called cooling rate $\zeta$ in the energy balance equation. The cooling rate (which vanishes for elastic collisions) measures the rate of energy dissipation due to inelastic collisions \cite{BP04,G19}.

One of the most intriguing and surprising effects of inelasticity in granular mixtures is the failure of energy equipartition in the homogeneous cooling state \cite{G19}. This means that the partial temperatures $T_i$ of each component are different from the global granular temperature. The energy nonequipartition (which is only due to the inelastic character of collisions) has important effects in problems such thermal diffusion segregation \cite{G08a,G09,G11}. On the other hand, a new contribution to the breakdown of energy equipartition (additional to the one caused by the inelasticity in collisions) has been reported very recently \cite{GGG19b}. Although considered in previous works of ordinary gases \cite{KS79a,KS79b,LCK83}, this new contribution  (which is associated with a nonzero first-order term $T_i^{(1)}$ in the expansion of the partial temperatures in powers of the gradients) had not been accounted in previous works of granular mixtures. Since $T_i^{(1)}$ is proportional to the divergence of the flow velocity $\mathbf{U}$ (i.e., $T_i^{(1)}=\varpi_i \nabla \cdot \mathbf{U}$), the coefficient $\varpi_i$ is also involved in the evaluation of the first-order contribution $\zeta^{(1)}$ to the cooling rate $\zeta$.

Although the coefficients $\varpi_i$ have been computed in the case of \emph{dry} granular mixtures (namely, a granular mixture where the influence of the interstitial fluid on the dynamics of grains is neglected), we are not aware of a similar calculation for binary granular suspensions at low density. The objective of this paper is to evaluate the coefficients $\varpi_i$ in granular suspensions where the effect of the surrounding gas is modeled by means of an effective external force \cite{KH01} composed by two terms: (i) a viscous drag term that mimics the friction of grains with the surrounding fluid and (ii) a stochastic term representing random and uncorrelated collisions between grains and fluid molecules. Once the first-order $T_i^{(1)}$ contributions to the partial temperatures are evaluated, as a complementary goal, we will also assess the impact of $T_i^{(1)}$ on $\zeta^{(1)}$.

In this work, we consider a binary granular suspension of smooth hard spheres in the low-density regime. In contrast to dry granular mixtures  \cite{GD02,GMD06,GM07}, the  first-order contributions to the partial temperatures are different from zero, even in the low-density regime. The determination of these quantities along with their influence on the cooling rate $\zeta^{(1)}$ are the main objectives of the present paper.

\section{Boltzmann Kinetic Equation for Polydisperse Gas-Solid Flows}
We consider a granular binary mixture of spheres of masses $m_i$ and diameters $\sigma_i$ ($i= 1,2$). Spheres are assumed to be completely smooth so that, the inelasticity of collisions between particles of component $i$ with $j$ is characterized by the constant (positive) coefficients of restitution $\alpha_{ij}\leq 1$. The solid particles are immersed in a molecular gas of viscosity $\eta_g$. In the low-density regime, the set of Boltzmann equations for the one-particle velocity distribution functions $f_i(\mathbf{r},\mathbf{v},t)$ of the component $i$ reads \cite{G19}
\begin{equation}
\label{2.1}
\frac{\partial f_i}{\partial t}+\mathbf{v}\cdot \nabla f_i-\gamma_i \Delta \mathbf{U}\cdot \frac{\partial f_i}{\partial\mathbf{v}}-\gamma_i\frac{\partial}{\partial\mathbf{v}}\cdot\mathbf{V}f_i-\frac{\gamma_i T_{\text{ex}}}{m_i}\frac{\partial^2 f_i}{\partial v^2}=\sum_{j=1}^2\; J_{ij}[f_i,f_j].
\end{equation}
Here, $\Delta\mathbf{U}=\mathbf{U}-\mathbf{U}_g$, $\mathbf{U}$ and $\mathbf{U}_g$ being the mean flow velocities of the solid particles and the interstitial gas, respectively, $\mathbf{V}=\mathbf{v}-\mathbf{U}$ is the peculiar velocity, and
\begin{equation}
\label{2.2}
J_{ij}[\mathbf{v}_1|f_i,f_j]=\sigma_{ij}^{d-1}\int\dd \mathbf{v}_2\int\dd\widehat{\boldsymbol{\sigma}}\;\Theta
\left(\widehat{\boldsymbol{\sigma}}\cdot\mathbf{g}_{12}\right)
\left(\widehat{\boldsymbol{\sigma}}\cdot\mathbf{g}_{12}\right)\Big[\alpha_{ij}^{-2}
f_i(\mathbf{v}_1'')
f_j(\mathbf{v}_2'')-f_i(\mathbf{v}_1)
f_j(\mathbf{v}_2)\Big]
\end{equation}
is the Boltzmann collision operator for collisions between particles of the component $i$ with particles of the component $j$. Here, $\sigma_{ij}=(\sigma_i+\sigma_j)/2$, $\widehat{\boldsymbol{\sigma}}$ is a unit vector directed along the line of centers from the sphere of the component $i$ to that of the component $j$, $\Theta$ is the Heaviside step function, and $\mathbf{g}_{12}=\mathbf{v}_1-\mathbf{v}_2$ is the relative velocity of the colliding pair. Moreover, the precollisional velocities $(\mathbf{v}_1'',\mathbf{v}_2'')$ that lead to $(\mathbf{v}_1,\mathbf{v}_2)$ are given by
\begin{equation}
\label{2.3}
\mathbf{v}_1''=\mathbf{v}_1-\mu_{ji}(1+\alpha_{ij}^{-1})(\widehat{\boldsymbol{\sigma}}\cdot\mathbf{g}_{12})\widehat{\boldsymbol{\sigma}},
\quad \mathbf{v}_2''=\mathbf{v}_2+\mu_{ij}(1+\alpha_{ij}^{-1})(\widehat{\boldsymbol{\sigma}}\cdot\mathbf{g}_{12})\widehat{\boldsymbol{\sigma}},
\end{equation}
where $\mu_{ij}=m_i/(m_i+m_j)$.

At low Reynolds numbers, it has been assumed in Eq.\ \eqref{2.1} that the effect of the surrounding molecular gas on the solid particles is modeled by a drag force proportional to $\mathbf{v}-\mathbf{U}_g$ plus a stochastic Langevin force representing Gaussian white noise. While the drag force mimics the friction of grains with the interstitial gas, the stochastic force models the kinetic energy gain of solid particles due to the interaction with the background gas. In addition, $\gamma_i$ is the drag or friction coefficient associated with the component $i$ and $T_{\text{ex}}$ is the temperature of the external gas. Note that the viscosity of the solvent $\eta_g \propto \sqrt{T_\text{ex}}$. More details of this kind of Langevin-like models can be found in Refs. \cite{HTG17,KIB14}. Here, for the sake of simplicity, we consider the coefficients $\gamma_i$ to be scalars proportional to the viscosity $\eta_g$ \cite{KH01}. In this case, according to the results obtained in lattice-Boltzmann simulations \cite{YS09b}, $\gamma_i$ is a function of the partial volume fractions
\beq
\label{2.3.1}
\phi_i=\frac{\pi^{d/2}}{2^{d-1}d\Gamma\left(\frac{d}{2}\right)}n_i\sigma_i^d,
\eeq
$n_i$ being the number density of the component $i$. Based on the restriction that in the dilute limit every particle is only subjected to its respective Stokes drag \cite{YS09b}, for hard spheres ($d=3$) and binary mixtures $(i=1,2)$, the friction coefficients $\gamma_i$ can be written as
\begin{equation}
\label{2.4}
\gamma_i=\gamma_0 R_i, \quad \gamma_0=\frac{18\eta_g}{\rho\sigma_{12}^2}, \quad R_i=\frac{\rho\sigma_{12}^2}{\rho_i\sigma_i^2}\phi_i,
\end{equation}
where $\rho=\sum_i \rho_i$ is the total mass density and $\rho_i=m_in_i$ is the mass density of the component $i$.

\section{Homogeneous Steady States}
Before analyzing inhomogeneous states, it is desirable to study first the homogeneous case. In this situation, the partial densities $n_i$ are constant, the granular temperature $T$ is spatially uniform, and with an appropriate selection of the reference frame, the mean flow velocities vanish $(\mathbf{U}=\mathbf{U}_g=\mathbf{0})$. After a transient regime, the system is expected to reach a \textit{steady} state and so, Eq.\ \eqref{2.1} becomes
\begin{equation}\label{3.1}
-\gamma_i\frac{\partial}{\partial\mathbf{v}}\cdot\mathbf{v}f_i-\frac{\gamma_i T_{\text{ex}}}{m_i}\frac{\partial^2 f_i}{\partial v^2}=\sum_{j=1}^2\; J_{ij}[f_i,f_j].
\end{equation}

For elastic collisions $(\alpha_{ij}=1)$, the cooling rate vanishes and in the case that $\gamma_1=\gamma_2= \gamma$, Eq.\ \eqref{3.1} admits the Maxwellian solution with a common temperature $T_1=T_2=T$. For inelastic collisions $(\alpha_{ij}\neq 1)$, $\zeta\ne 0$ and to date the solution of Eq.\ \eqref{3.1} is not known. Thus, one has to consider approximate forms for $f_i$. Here, for the sake of simplicity and to compute the first velocity moments of $f_i$, we will replace $f_i$ by the Maxwellian distribution at the temperature $T_i$:
\begin{equation}
\label{3.2}
f_i(\mathbf{v})\rightarrow f_{i,\text{M}}(\mathbf{v})=n_i\left(\frac{m_i}{2\pi T_i}\right)^{d/2}\text{exp}\left(-\frac{m_iv^2}{2 T_i}\right),
\end{equation}
where the partial temperatures $T_i$ are defined as
\begin{equation}
\label{3.3}
 T_i=\frac{m_i}{dn_i}\int\dd\mathbf{v}\;V^2f_i(\mathbf{v}).
\end{equation}

\begin{figure}[!h]
  \centering
  \includegraphics[width=0.45\textwidth]{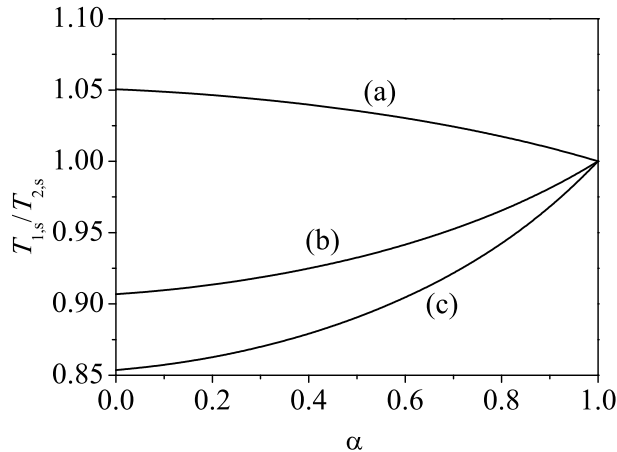}
  \caption{Plot of the temperature ratio $T_{1,\text{s}}/T_{2,\text{s}}$ in the homogeneous steady state as a function of the common coefficient of restitution $\alpha$ for an equimolar mixture ($x_1=\frac{1}{2}$) of hard spheres ($d=3$) with $\sigma_1/\sigma_2=1$, and $T_\text{ex}^*=0.1$. Three different values of the mass ratio are considered: $m_1/m_2=0.5$ (a), $m_1/m_2=4$ (b), and $m_1/m_2=10$ (c).}
	\label{fig1}
\end{figure}

The (reduced) partial temperatures $\tau_{i,\text{s}}=T_{i,\text{s}}/T_{\text{s}}$ can be obtained by multiplying both sides of Eq.\ \eqref{3.1} by $m_iv^2$ and integrating over velocity. The result is
\begin{equation}
\label{3.4}
2\gamma^*_{i,\text{s}}(\tau_{i,\text{s}}-\theta_\text{s}^{-1})+\zeta_{i,\text{s}}^*\tau_{i,\text{s}}=0,
\end{equation}
where the subscript $\text{s}$ means that all the quantities are evaluated in the steady state. In Eq.\ \eqref{3.4},
$\gamma_{i,\text{s}}^*=\ell\gamma_{i,\text{s}}/v_{0\text{s}}$, $\theta_{\text{s}}=T_{\text{s}}/T_{\text{ex}}$, $\zeta_{i,\text{s}}^*=\ell\zeta_{i,\text{s}}/v_{0\text{s}}$, $v_{0\text{s}}=\sqrt{2T_{\text{s}}/\overline{m}}$ is the thermal speed, $\overline{m}=\sum_im_i/2$, and $\ell=1/n\sigma^{d-1}_{12}$ is the mean free path for elastic hard spheres. Here,
\begin{equation}
\label{3.5}
\zeta_{i,\text{s}}=-\frac{m_i}{dn_iT_{i,\text{s}}}\sum_{j=1}^2\int\dd\mathbf{v}\; V^2J_{ij}[f_{i,\text{s}},f_{j,\text{s}}]
\end{equation}
are the partial cooling rates for the partial temperatures $T_{i,\text{s}}$. They can be estimated by considering the Maxwellian approximation \eqref{3.2} for $f_{i,\text{s}}$. In reduced form, the result for the partial cooling rates is
\begin{equation}
\label{3.6}
\zeta^*_{i,\text{s}}=\frac{4\pi^{\left(d-1\right)/2}}{d\Gamma\left(\frac{d}{2}\right)}\sum_{j=1}^2\;x_{j}\mu_{ji}
\left(\frac{\sigma_{ij}}{\sigma_{12}}\right)^{d-1}\left(\frac{\beta_i+\beta_j}{\beta_i\beta_j}\right)^{1/2}
\left(1+\alpha_{ij}\right)\left[1-\frac{\mu_{ji}}{2}\left(1+\alpha_{ij}\right)
\frac{\beta_i+\beta_j}{\beta_j}\right],
\end{equation}
where $\beta_i=m_iT_{\text{s}}/\overline{m}T_{i,\text{s}}$. In contrast to the dry granular case \cite{GD99b}, the (reduced) partial temperatures $\tau_{i,\text{s}}=T_{i,\text{s}}/T_{\text{s}}$ depend on the granular temperature through the (dimensionless) drag coefficients $\gamma_{i,\text{s}}^*$. This dependence can be made more explicit when one express $\gamma_{i,\text{s}}^*$ in terms of the mole fraction $x_1=n_1/(n_1+n_2)$, the volume fractions $\phi_i$, and the (reduced) temperature $\theta_{\text{s}}$ as
\begin{equation}
\label{3.7}
\gamma_{i,\text{s}}^*=\lambda_i\theta_{\text{s}}^{-1/2},\quad
\lambda_i=\frac{\sqrt{2}\pi^{d/2}}{2^dd\Gamma\left(\frac{d}{2}\right)}\frac{R_i}
{\sqrt{T_{\text{ex}}^*}\sum_j(\sigma_{12}/\sigma_j)^d\phi_j}.
\end{equation}
Here, $T_{\text{ex}}^*\equiv T_{\text{ex}}/\left(\overline{m}\sigma_{12}^{2}\gamma_0^2\right)$ is the (dimensionless) background gas temperature. Substitution of Eq.\ \eqref{3.6} into the coupled equations \eqref{3.4} allows us to obtain the partial temperatures in terms of the concentration $x_1$, the volume fractions $\phi_i$, the (reduced) background temperature $T_{\text{ex}}^*$, and the parameters of the mixture (masses, diameters, and coefficients of restitution).

The temperature ratio $T_{1,\text{s}}/T_{2,\text{s}}$ is plotted in Fig.\ \ref{fig1} versus the (common) coefficient of restitution $\alpha\equiv \alpha_{11}=\al_{22}=\al_{12}$ for $x_1=\frac{1}{2}$, $\sigma_1=\sigma_2$, $T_\text{ex}^*\equiv T_{\text{ex}}/\left(\overline{m}\sigma_{12}^{2}\gamma_0^2\right)=0.1$, and three different values of the mass ratio. As expected, it is quite apparent that the extent of the energy violation is greater when the mass disparity is large.

\section{Chapman-Enskog Solution of the Boltzmann Equation}
We now slightly perturb the homogeneous steady state by small spatial gradients. These gradients give rise to nonzero contributions to the mass, momentum, and heat fluxes. Here, we want to compute the first-order contributions to the partial temperatures. In order to achieve them, we have to solve the Boltzmann equation \eqref{2.1} by means of the Chapman--Enskog method \cite{CC70} adapted to dissipative dynamics. As usual, the Chapman--Enskog method assumes the existence of a \textit{normal} or hydrodynamic solution where all the space and time dependence of the velocity distribution functions $f_i(\mathbf{v},\mathbf{r},t)$ occurs via a functional dependence on the hydrodynamic fields. For small spatial gradients, this functional dependence can be made local in space through an expansion of the distribution functions in powers of the spatial gradients: $f_i\rightarrow f_i^{(0)}+f_i^{(1)}+\cdots$. Here, only terms up to first order in gradients will be retained (Navier--Stokes hydrodynamic order).

Moreover, in ordering the different levels of approximation in the Boltzmann equation, one has to characterize the magnitude of the friction coefficients $\gamma_i$ and the difference of the mean velocities $\Delta\mathbf{U}$ relative to the spatial gradients. With respect to $\gamma_i$, since these coefficients do not induce any flux in the suspension they are considered to be to zeroth-order in gradients. In the case of $\Delta\mathbf{U}$, the fact that the mean flow velocity $\mathbf{U}$ relaxes to $\mathbf{U}_g$ in the homogeneous state implies that $\Delta\mathbf{U}$ is considered to be at least of first order in the perturbation scheme.

\subsection{Zeroth-Order Approximation}
To zeroth-order in the perturbation expansion, the Boltzmann equation \eqref{2.1} is given by
\begin{equation}
\label{4.1.1}
\partial_t^{(0)}\partial f^{(0)}_i-\gamma_i\frac{\partial}{\partial\mathbf{v}}\cdot\mathbf{V}f_i^{(0)}-\frac{\gamma_i T_{\text{ex}}}{m_i}\frac{\partial^2 f_i^{(0)}}{\partial v^2}=\sum_{j=1}^2\; J_{ij}[f_i^{(0)},f_j^{(0)}].
\end{equation}
The balance equation at this order give $\partial_t^{(0)}n_i=0$, $\partial_t^{(0)}\mathbf{U}=\mathbf{0}$, and $\partial_t^{(0)}T=\Lambda^{(0)}$. Here,
\begin{equation}\label{4.1.2}
\Lambda^{(0)}\equiv2\sum_{i=1}^2x_i\gamma_i\left(\theta^{-1}-\tau_i\right)-\zeta^{(0)},
\end{equation}
where $\zeta^{(0)}=\sum_i x_i\tau_iv_0\zeta^*_{i,0}/\ell$ is determined by Eq.\ \eqref{3.6} to zeroth-order.

The kinetic equation for the temperature ratio $\tau_i=T_i^{(0)}/T$ (where $T_i^{(0)}$ denotes the zeroth-order contribution to the partial temperature $T_i$) can be easily obtained from Eq.\ \eqref{4.1.1} with the result
\begin{equation}
\label{4.1.3}
\Lambda^* \theta \frac{\partial \tau_i}{\partial \theta}=-\tau_i \Lambda^*+\Lambda_i^*,
\end{equation}
where
\begin{equation}
\label{4.1.4}
\Lambda^*=\frac{\ell \Lambda^{(0)}}{v_0}=x_1\Lambda_1^*+x_2\Lambda_2^*,
\end{equation}
and
\begin{equation}
\label{4.1.5}
\Lambda_i^*=2\gamma_i^*\left(\theta^{-1}-\tau_i \right)-\tau_i \zeta_{i,0}^*.
\end{equation}
Here, $\zeta_{i,0}^*=\ell \zeta_i^{(0)}/v_0$ where $\zeta_i^{(0)}$ is defined by Eq.\ \eqref{3.6}.

In the steady state, $\Lambda^*=\Lambda_i^*=0$ and Eqs.\ \eqref{4.1.3} are consistent with Eqs.\ \eqref{3.4} for $i=1,2$. To evaluate the first-order contributions to the partial temperatures $T_i^{(1)}$ in the steady state one needs to know the derivatives $\Delta_{\theta,1}\equiv(\partial\tau_1/\partial\theta)_{\text{s}}$ and $\Delta_{\lambda_1,1}\equiv(\partial\tau_1/\partial\lambda_1)_{\text{s}}$. Since $\lambda_2=(R_2/R_1)\lambda_1$, then $(\partial\tau_2/\partial\lambda_1)_{\text{s}}=(R_1/R_2)(\partial\tau_1/\partial\lambda_1)_{\text{s}}$. Analytical expressions of these derivatives can be obtained in the vicinity of the steady state. The forms of $\Delta_{\theta,1}$ and $\Delta_{\lambda_1,1}$ are displayed in the Appendix \ref{appA}.

\subsection{First-Order Contributions to the Partial Temperatures}
The first-order velocity distribution functions $f_i^{(i)}$ obey the following kinetic equation
\beqa
\label{4.2.1}
\partial_t^{(0)}f_i^{(1)}-\gamma_i\frac{\partial}{\partial\mathbf{v}}\cdot\mathbf{V}f_i^{(1)}&-&\frac{\gamma_i T_{\text{ex}}}{m_i}\frac{\partial^2 f_i^{(1)}}{\partial v^2}= -\left(D_t^{(1)}+\mathbf{V}\cdot\nabla\right)f_i^{(0)}\nonumber\\
& +&\gamma_i\Delta\mathbf{U}\cdot\frac{\partial f_i^{(0)} }{\partial\mathbf{v}}+\sum_{j=1}^2\; \left(J_{ij}\left[f_i^{(1)},f_j^{(0)}\right]+J_{ij}\left[f_i^{(0)},f_j^{(1)}\right]\right),
\eeqa
where $D_t^{(1)}\equiv\partial_t^{(1)}+\mathbf{U}\cdot\nabla$. To first order, the balance equations are
\begin{equation}\label{4.2.2}
D_t^{(1)}n_i=-n_i\nabla\cdot\mathbf{U},\quad D_t^{(1)}\mathbf{U}=-\rho^{-1}\nabla p-\Delta\mathbf{U}\sum_{i=1}^2\frac{\rho_i}{\rho}\gamma_i+\rho^{-1}\left(\gamma_1-\gamma_2\right)\mathbf{j}_1^{(1)},
\end{equation}
\begin{equation}\label{4.2.3}
D_t^{(1)}T=-\frac{2p}{dn}\nabla\cdot\mathbf{U}-\zeta^{(1)}T-2\sum_{i=1}^2\gamma_i x_iT_i^{(1)}.
\end{equation}
Here, $p=nT$ is the hydrostatic pressure and $\mathbf{j}_1^{(1)}$ is the mass flux to first order in spatial gradients. Since $T_i^{(1)}$ and $\zeta^{(1)}$ are scalars, they are coupled to $\nabla\cdot\mathbf{U}$ and hence, they can be written as $T_i^{(1)}=\varpi_i\nabla\cdot\mathbf{U}$ and $\zeta^{(1)}=\zeta_U\nabla\cdot\mathbf{U}$. Note that $n_1T_1^{(1)}+n_2T_2^{(1)}=0$ and so, $\varpi_2=-(x_1/x_2)\varpi_1$.


We compute now the first-order contributions $T_i^{(1)}$ to the partial temperatures. They are defined as
\begin{equation}\label{4.2.4}
T_i^{(1)}=\frac{m_i}{dn_i}\int\dd\mathbf{v}\;V^2\;f_i^{(1)}(\mathbf{V}).
\end{equation}
To determine the coefficients $\varpi_i$, we multiply both sides of Eq.\ \eqref{4.2.1} by $m_iV^2$ and integrate over velocity. In the leading Sonine approximation, one gets the set of algebraic equations
\begin{equation}
\label{4.2.5}
\sum_{j=1}^2\Big[\omega_{ij}+2\gamma_j x_j\left(\tau_i+\theta \Delta_{\theta,i}\right)-2\gamma_i\delta_{ij}+\left(T_i^{(0)}+T\theta \Delta_{\theta,i}\right)\xi_j\Big]\varpi_j
=-\frac{2}{d}T\theta\Delta_{\theta,i}
-T\sum_{j=1}^2n_j\frac{\partial \lambda_1}{\partial n_j}\Delta_{\lambda_1,i}.
\end{equation}
Here,
\beq
\label{4.2.5.1}
n_j\frac{\partial \lambda_1}{\partial n_j}=\frac{(m_1-m_2)n_j}{\rho}\Big(x_2 \delta_{1j}-x_1 \delta_{2j}\Big)\lambda_1,
\eeq
and upon writing Eq.\ \eqref{4.2.5} use has been made of the relation
\begin{equation}\label{4.2.6}
\zeta_U=\sum_{i=1}^2\xi_i\varpi_i,
\end{equation}
where the coefficients $\xi_i$ are
\begin{equation}
\label{4.2.7}
\xi_i=\frac{3\pi^{(d-1)/2}}{2d\Gamma\left(\frac{d}{2}\right)}\frac{v_0^3m_i}{n T T_i^{(0)}}\sum_{j=1}^2 n_i n_j \sigma_{ij}^{d-1}\mu_{ji}(1-\alpha_{ij}^2)\left(\beta_i+\beta_j\right)^{1/2}\beta_i^{-3/2}\beta_j^{-1/2}.
\end{equation}
In the Maxwellian approximation \eqref{3.2}, the collision frequencies $\omega_{ii}$ and $\omega_{ij}$ are given by \cite{GGG19b}
\beqa
\label{4.2.8}
\omega_{ii}&=&-\frac{\pi^{(d-1)/2}}{2dT_i^{(0)}\Gamma\left(\frac{d}{2}\right)}v_0^3\Bigg\{\frac{3}{\sqrt{2}}
n_i\sigma_i^{d-1}m_i \beta_i^{-3/2}
\left(1-\alpha_{ii}^2\right)-n_j m_{ij} \sigma_{ij}^{d-1}\left(1+\alpha_{ij}\right)\left(\beta_i+\beta_j\right)^{-1/2}\beta_i^{-3/2}\beta_j^{-1/2}\nonumber\\
& & \times
\Big[3\mu_{ji}\left(1+\alpha_{ij}\right)\left(\beta_i+\beta_j\right)-2\left(2\beta_i+3\beta_j\right)\Big]\Bigg\},
\eeqa
\begin{equation}
\label{4.2.9}
\omega_{ij}=\frac{\pi^{(d-1)/2}}{2dT_j^{(0)}\Gamma\left(\frac{d}{2}\right)}v_0^3n_jm_{i}\mu_{ij}\sigma_{ij}^{d-1}
\left(1+\alpha_{ij}\right)\left(\beta_i+\beta_j\right)^{-1/2}\beta_i^{-1/2}\beta_j^{-3/2}\Big[3\mu_{ji}\left(1+\alpha_{ij}\right)
\left(\beta_i+\beta_j\right)-2\beta_j\Big].
\end{equation}
In Eqs.\ \eqref{4.2.8}--\eqref{4.2.9}, it is understood that $i\ne j$.

\begin{figure}[!h]
  \centering
  \includegraphics[width=0.45\textwidth]{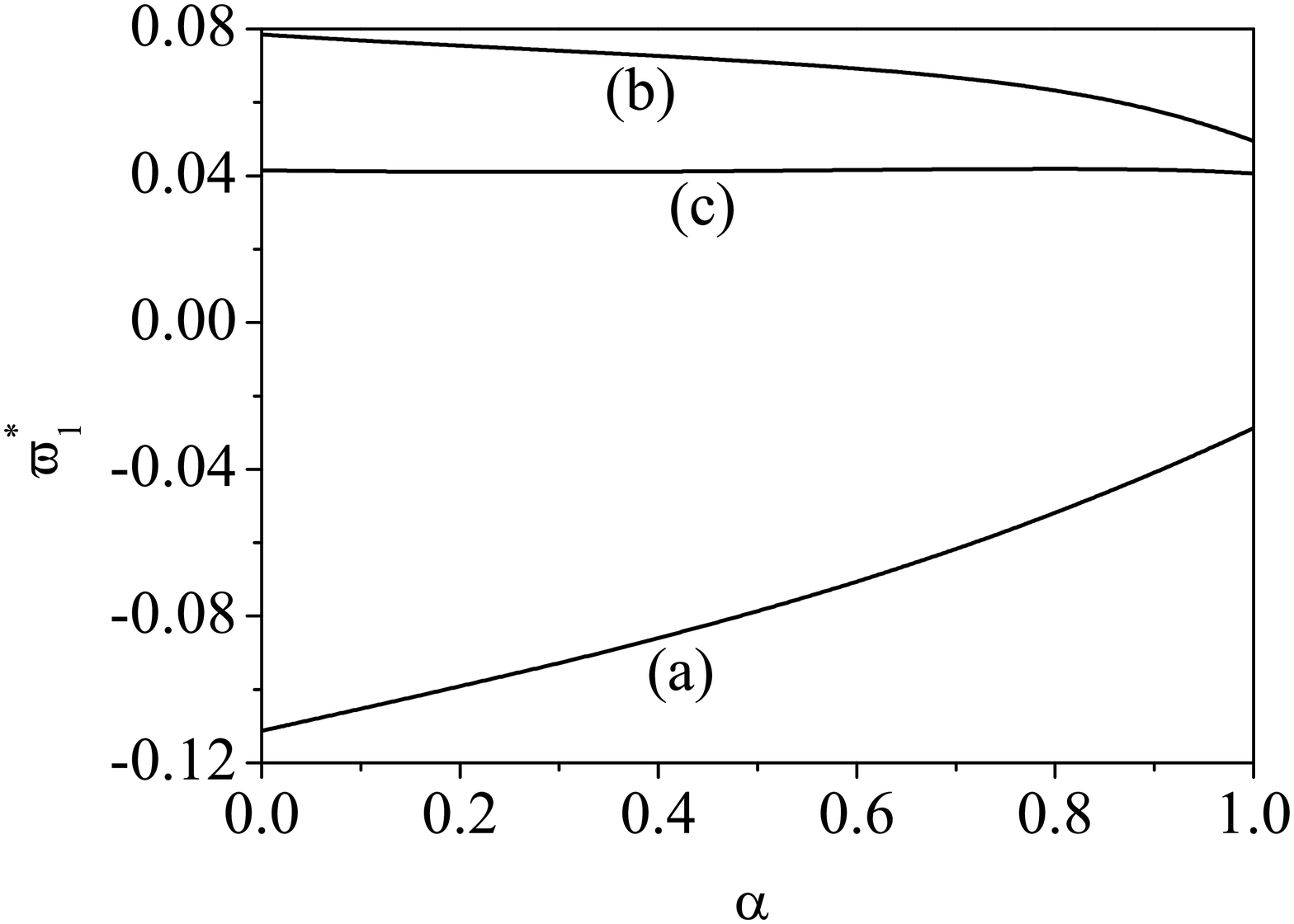}
  \includegraphics[width=0.45\textwidth]{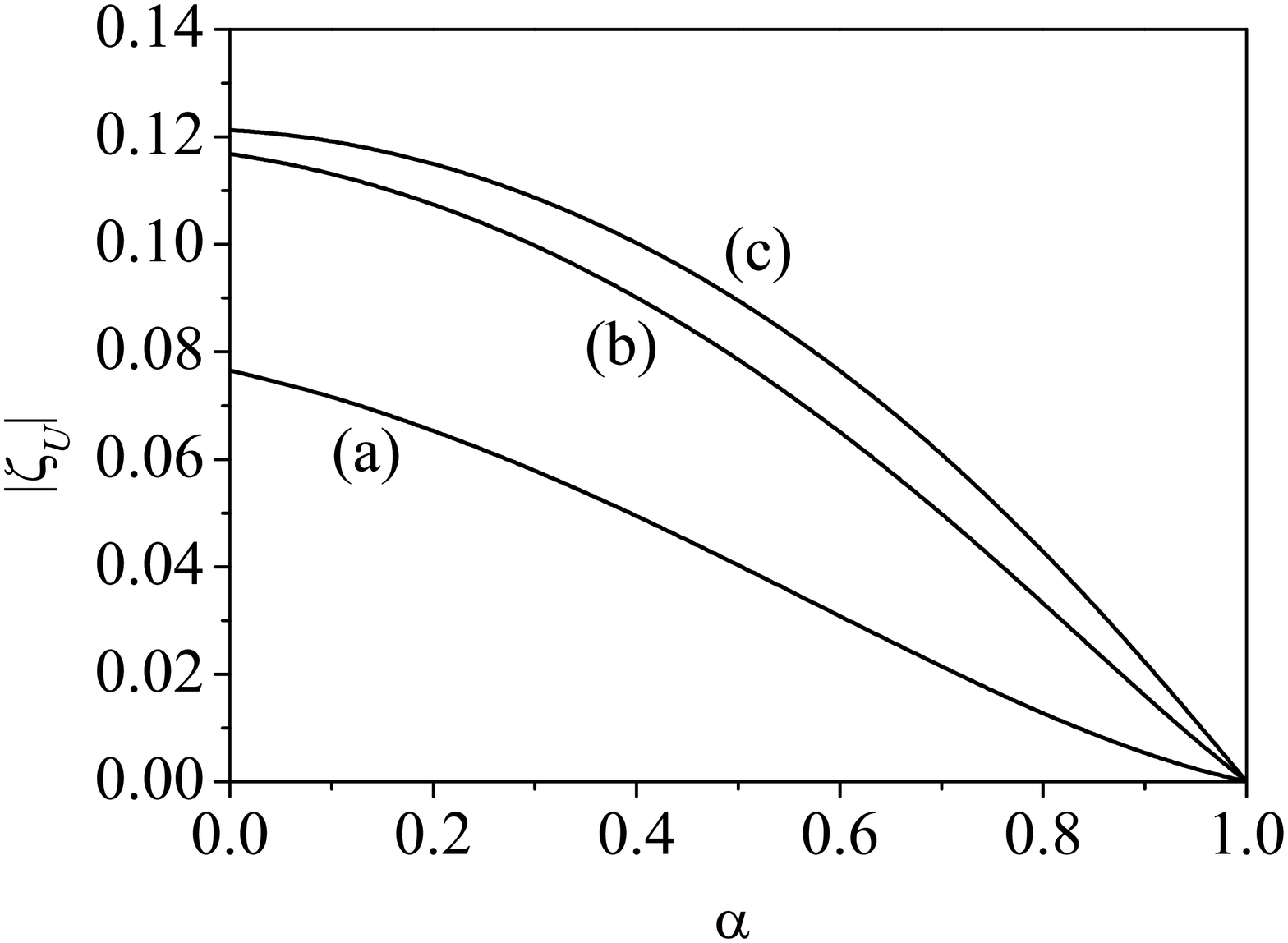}
  \caption{Plot of the (reduced) coefficients $\varpi_1^*$ and $\zeta_U$ as a function of the common coefficient of restitution $\alpha$ for an equimolar mixture ($x_1=\frac{1}{2}$) of hard spheres ($d=3$) with $\sigma_1/\sigma_2=1$, and $T_\text{ex}^*=0.1$. Three different values of the mass ratio are considered: $m_1/m_2=0.5$ (a), $m_1/m_2=4$ (b), and $m_1/m_2=10$ (c).}
	\label{fig2}
\end{figure}

The solution to the set of equations \eqref{4.2.5} provides the explicit forms of $\varpi_1$ and $\varpi_2$. It is seen that $\varpi_2=-(x_1/x_2)\varpi_1$, as the solubility conditions of the Chapman--Enskog method requires. Figure \ref{fig2} shows the $\alpha$-dependence of $\varpi_1^*\equiv (n\sigma_{12}^2v_0/T)\varpi_1$ and $\zeta_U$ for different systems. Here, $n=n_1+n_2$ is the total number density. We observe that the influence of the inelasticity on both $\varpi_1^*$ and $\zeta_U$ is important, specially for strong inelasticity. Thus, both quantities should be taken into account in the kinetic description of binary granular suspensions.

In summary, we have determined the first-order contributions $T_i^{(1)}$ to the partial temperatures in binary granular suspensions. The fact that $T_i^{(1)}\neq 0$ yields a \textit{new} contribution (additional to the one caused by inelasticity in collisions) to the breakdown of energy equipartition. Since this contribution is proportional to the divergence of the flow velocity,  it is involved  then in the evaluation of the first-order contribution $\zeta_U$ to the cooling rate. Our results show that the magnitude of both coefficients $T_i^{(1)}$ and $\zeta_U$ can be significant in some regions of the parameter space of the system. This conclusion contrasts with the results obtained for dry granular mixtures \cite{GD02,GMD06,GM07} where it was shown that both coefficients vanish in the low density limit.

\acknowledgments

The present work has been supported
by the Spanish Government through Grant No. FIS2016-
76359-P and by the Junta de Extremadura (Spain) Grants
No. IB16013 (V.G.) and No. GR18079, partially financed by
``Fondo Europeo de Desarrollo Regional'' funds. The research
of R.G.G. has been supported by the predoctoral fellowship
BES-2017-079725 from the Spanish Government.

\appendix

\section{Derivatives of the temperature ratio near the steady state \label{appA}}

In this Appendix, we determine the derivatives of the temperature ratio $\tau_1=T_1^{(0)}/T$ with respect to $\theta$ and $\lambda_1$ near the steady state. We consider first the derivative $(\partial \tau_1/\partial \theta)_{x_1,\lambda_1}$. To achieve it, we consider Eq.\ \eqref{4.1.3} when $i=1$:
\beq
\label{a1}
\Lambda^* \theta \frac{\partial \tau_1}{\partial \theta}=-\tau_1 \Lambda^*+\Lambda_1^*,
\eeq
where $\Lambda^*$ and $\Lambda_1^*$ are defined by Eqs.\ \eqref{4.1.4} and \eqref{4.1.5}, respectively. According to Eq.\ \eqref{3.6}, the (reduced) partial cooling rate $\zeta_{1,0}^*\equiv \ell \zeta_1^{(0)}/v_0$ can be written as
\beq
\label{a2}
\zeta_{1,0}^*=\tau_1^{1/2}M_1^{-1/2}\zeta_1'(x_1,\beta),
\eeq
where $\beta=\beta_1/\beta_2=m_1\tau_2/(m_2\tau_1)$, $\tau_2=(1-x_1\tau_1)/x_2$, and
\beqa
\label{a3}
\zeta_1'(x_1,\beta)&=&\frac{\sqrt{2}\pi^{\left(d-1\right)/2}}{d\Gamma\left(\frac{d}{2}\right)}x_1
\left(\frac{\sigma_{1}}{\sigma_{12}}\right)^{d-1}
\left(1-\alpha_{11}^2\right)+\frac{4\pi^{\left(d-1\right)/2}}{d\Gamma\left(\frac{d}{2}\right)}x_2\mu_{21}\left(1+\beta\right)^{1/2}
\left(1+\alpha_{12}\right)\nonumber\\
& & \times \left[1-\frac{\mu_{21}}{2}\left(1+\alpha_{12}\right)\left(1+\beta\right)\right].
\eeqa
At the steady state, $\Lambda^*=\Lambda_1^*=\Lambda_2^*=0$ and hence, according to Eq.\ \eqref{a1}, the derivative $\partial \tau_1/\partial \theta$ becomes indeterminate. On the other hand, as for dilute multicomponent granular suspensions \cite{KG13}, the above problem can be fixed by applying l'H\^opital's rule. In this case, we take first the derivative with respect to $\theta$ in both sides of Eq. \eqref{a1} and then take the steady-state limit. After some algebra, one easily obtains the following quadratic equation for  the derivative $\Delta_{\theta,1}=\left(\partial\tau_1/\partial\theta\right)_{\text{s}}$:
\beq
\label{a4}
\theta\Lambda_{1}^{(\theta)}\Delta_{\theta,1}^2+\left(\theta\Lambda_{0}^{(\theta)}+
\tau_{1}\Lambda_{1}^{(\theta)}-\Lambda_{11}^{(\theta)}\right)\Delta_{\theta,1}-
\Lambda_{10}^{(\theta)}\nonumber+\tau_{1}\Lambda_{0}^{(\theta)}=0,
\eeq
where $\Lambda_{0}^{(\theta)}=x_1 \Lambda_{10}^{(\theta)}+x_2\Lambda_{20}^{(\theta)}$ and $\Lambda_{1}^{(\theta)}=x_1 \Lambda_{11}^{(\theta)}+x_2\Lambda_{21}^{(\theta)}$. Here, we have introduced the quantities
\beq
\label{a5}
\Lambda_{10}^{(\theta)}=\gamma_1^*\theta^{-1}\tau_1-3\gamma_1^*\theta^{-2}, \quad \Lambda_{20}^{(\theta)}=\gamma_2^*\theta^{-1}\tau_2-3\gamma_2^*\theta^{-2},
\eeq
\beq
\label{a6}
\Lambda_{11}^{(\theta)}=-2\gamma_1^*-\frac{3}{2}\zeta_{10}^*+\tau_1^{-1/2}\frac{M_1^{1/2}}{x_2M_2}
\left(\frac{\partial\zeta'_1}{\partial\beta}\right)_{x_1}, \quad \Lambda_{21}^{(\theta)}=2\frac{x_1}{x_2}\gamma_2^*+\frac{3}{2}\frac{x_1}{x_2}\zeta_{20}^*+
\frac{M_1}{x_2M_2^{3/2}}\frac{\tau_2^{3/2}}{\tau_1^{2}}\left(\frac{\partial\zeta_2'}{\partial\beta}\right)_{x_1},
\eeq
where $\theta=T/T_\text{ex}$.  In Eqs.\ \eqref{a4}--\eqref{a6}, it is understood that all the quantities are evaluated in the steady state. As for dilute driven granular mixtures \cite{KG13}, an analysis of the solutions to Eq. \eqref{a4} shows that in general only one of the roots leads to a physical behavior of the diffusion coefficients. We take this root as the physical root of the quadratic equation \eqref{a4}.

Once the derivative $\Delta_{\theta,1}$ is known, we can determine the derivative $(\partial\tau_1/\partial\lambda_1)_{\theta,x_1}$ in a similar way. In order to get it, we take first the derivative of Eq.\ \eqref{a1} with respect to $\lambda_1$ and then consider the steady-state conditions. The final expression is
\beq
\label{a8}
\Delta_{\lambda_1,1}=-\frac{\tau_{1}\Lambda_{0}^{(\lambda_1)}-
\Lambda_{10}^{(\lambda_1)}+\theta \Lambda_{0}^{(\lambda_1)}\Delta_{\theta,1}}{\theta
\Lambda_{1}^{(\theta)}\Delta_{\theta,1}+\tau_{1}\Lambda_{1}^{(\theta)}
-\Lambda_{11}^{(\theta)}},
\eeq
where $\Lambda_{0}^{(\lambda)}=x_{1}\Lambda_{10}^{(\lambda_1)}+x_{2}\Lambda_{20}^{(\lambda_1)}$, and
\beq
\label{a9}
\Lambda_{10}^{(\lambda_1)}=2\theta^{-1/2}\left(\theta^{-1}-\tau_1\right),\quad \Lambda_{20}^{(\lambda)}=2\frac{R_2}{R_1}\theta^{-1/2}\left(\theta^{-1}-\tau_2\right).
\eeq


%
%

\end{document}